\documentclass{EWCG06}
\usepackage{url}
\usepackage{nicefrac}

\def\abs#1{\mathopen| #1 \mathclose|}

\def\ceil#1{\lceil #1 \rceil}
\def\paren#1{\ensuremath{\left({#1}\right)}}
\def\root#1{\ensuremath{\smash[b]{\sqrt{#1}}}}

\def\thePlane{\ensuremath{\mathbb{E}^2}}
\def\Surf{\ensuremath{\Sigma}}
\def\Pants{\ensuremath{\Pi}}

\def\e{\ensuremath{\varepsilon}}

\begin{document}
\title{Pants Decomposition of the Punctured Plane}

\author{Sheung-Hung Poon\thanks{Dept.\@ of Math and Computer Sci.,
Technische Universiteit Eindhoven, The Netherlands;
\{spoon,sthite\}@win.tue.nl.\qquad
{S.-H.P.}\@ and {S.T.}\@ were supported by the Netherlands'
Organisation for Scientific Research (NWO) under project
numbers~612.065.307 and~639.023.301 respectively.}
\and Shripad Thite\footnotemark[1]}

\index{Poon, Sheung-Hung}
\index{Thite, Shripad}


\maketitle

\begin{abstract}
A \emph{pants decomposition} of an orientable surface~$\Surf$ is a
collection of simple cycles that partition~$\Surf$ into
\emph{pants}, i.e., surfaces of genus zero with three boundary cycles.
Given a set~$P$ of~$n$ points in the plane~$\thePlane$, we consider
the problem of computing a pants decomposition of $\Surf = \thePlane
\setminus P$ of minimum total length.  We give a
polynomial-time approximation scheme using Mitchell's guillotine
rectilinear subdivisions.  We give an $O(n^4)$-time algorithm to
compute the shortest pants decomposition of~$\Surf$ when the cycles
are restricted to be axis-aligned boxes, and an $O(n^2)$-time algorithm when all the
points lie on a line; both exact algorithms use
dynamic programming with Yao's speedup.
\end{abstract}

\section{Introduction}

\emph{Surfaces} (\emph{$2$-manifolds}), such as spheres,
cylinders, tori, and more, are commonly encountered topological spaces in
applications like computer graphics and geometric modeling.
To understand the topology of the surface or to compute various properties of the surface, it is useful to \emph{decompose} the surface into simple
parts.  Among the possible ways to decompose a
given surface, it is desirable to compute an \emph{optimum}
decomposition, one that minimizes a metric depending on the
application.

A decomposition of an orientable surface~$\Surf$ that has been studied is a \emph{pants decomposition}~\cite{hatcher99pants,colindeverdiere03optimal}, a collection of disjoint cycles that
partition~$\Surf$ into \emph{pants}, where a \emph{pant}\footnotemark{} is a
surface of genus zero with three boundary cycles. Every compact
orientable surface---except the sphere, disk, cylinder, and
torus---admits a pants decomposition~\cite{colindeverdiere03optimal}.

A natural measure of a pants decomposition to minimize is the total length of its boundary cycles. The \emph{length} of a pants decomposition~$\Pants$ of~$\Surf$, denoted by~$\abs{\Pi}$, is the sum of the (Euclidean) lengths of all the cycles in~$\Pants$. (If a
subsegment is traversed more than once, its length is counted with multiplicity.)  A \emph{non-crossing pants decomposition} is a pants decomposition
that allows any two cycles to touch as long as
they do not cross transversely.  A \emph{shortest} pants decomposition
is a non-crossing pants decomposition of minimum length.

\begin{figure}[hb]\centering\small
\includegraphics[height=4cm]{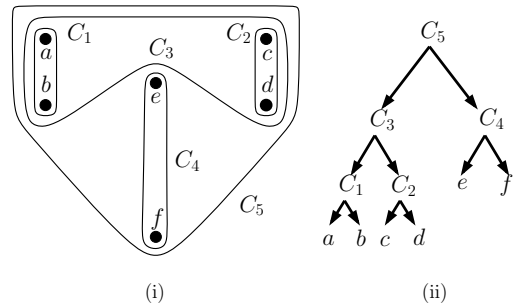}
\caption{\small\sf (i)~A pants decomposition of the plane punctured by a set of~$6$
points, and (ii)~the corresponding binary tree.}
\label{fig:pants}
\end{figure}

The problem of computing a shortest pants decomposition of an arbitrary surface~$\Surf$ is open.  In this paper, we study a variant of the problem where~$\Surf$ is the \emph{punctured plane}, i.e., $\Surf = \thePlane \setminus P$ where $P$ is a discrete set of~$n$ points.  Figure~\ref{fig:pants}(i) gives an example pants decomposition of the plane with 6 punctures.
Colin de Verdi\`ere and Lazarus~\cite{colindeverdiere03optimal} studied a related problem: given a pants decomposition of an arbitrary surface, they compute a \emph{homotopic} pants decomposition in which each cycle is a shortest cycle in its homotopy class.

\footnotetext{We call a surface of genus
zero with three boundary components a \emph{pant} instead of a
\emph{pair of pants}~\cite{colindeverdiere03optimal}.  We
refer to two pants instead of two pairs of pants; the latter phrase
can be misunderstood to mean four such surfaces.}

A cycle~$C$ on~$\Surf$ is
\emph{essential} if it does not bound a disk or an annulus.  A pants
decomposition, also called a \emph{maximal cut
system}~\cite{hatcher99pants}, is naturally obtained by the following
greedy procedure. Let~$C$ be an essential cycle. Cut~$\Surf$ along~$C$
to get a surface~$\Surf'$ with two additional boundary cycles $C_1$
and~$C_2$ corresponding to the two sides of the cut.  The cycles $C_1$
and~$C_2$, together with the set of cycles obtained by recursing
on~$\Surf'$, is a pants decomposition~$\Pants$ of~$\Surf$. The resulting cycle structure can be modeled as a binary tree with $n$ leaves (Figure~\ref{fig:pants}(ii)).
Each cycle~$C$ of~$\Pants$ is essential because it encloses two other
cycles, say~$C_1$ and~$C_2$; we write \emph{$C_1 \prec C$} and
\emph{$C_2 \prec C$} to indicate that~$C$ encloses~$C_1$ and~$C_2$. We
say that two cycles~$C_1$ and~$C_2$ are \emph{independent} if neither
$C_1 \prec C_2$ nor $C_2 \prec C_1$.  The final result is a pants decomposition of a bounded subset of the plane together with a single unbounded component.

In this paper, we give a simple algorithm with approximation ratio $O(\log n)$, and a polynomial-time approximation scheme (PTAS) using Mitchell's guillotine
rectilinear subdivisions.  We compute the shortest pants decomposition in $O(n^4)$ time when the cycles are restricted to be axis-aligned boxes, and in $O(n^2)$ time when all the points lie on a line; both exact algorithms use dynamic programming with Yao's speedup, and are faster by a linear factor than the `na\"ive' dynamic programming formulations.

\section{A simple approximation algorithm}
\label{sec:logapprox}

If~$\Pants^*$
is a shortest pants decomposition of the punctured plane~$\Surf=\thePlane \setminus P$, then every cycle in~$\Pants$ is a simple polygon whose vertices belong to~$P$.
We argue next that~$\Pants^*$ contains a traveling salesperson (TSP) tour of the
points.  Choose any vertex on the outermost cycle as the start of the
tour. Traverse the outermost cycle counterclockwise. The first time we
visit a vertex~$u$ that also belongs to an inner cycle (one of the two
legs of the pant) that has not been traversed yet, we recursively
construct a tour beginning and ending at~$u$ that traverses the
unvisited vertices on or in the interior of this inner cycle. After
the recursion, we continue along the outermost cycle, repeating the
recursive traversal of the second leg of the pant, until we reach our
original starting point.
Hence, $\Pants^*$ must be at least as
long as a shortest Euclidean TSP tour~$T^*$ of the points. Hence,
$\abs{\Pants^*} \ge \abs{T^*}$.

We convert a TSP tour~$T$ of the points in~$P$ to a pants
decomposition~$\Pants$ as follows. Initially, $\Pants$ is the empty
set. Order the points from~$0$ through $n-1$ in the order along the
tour~$T$. Let $C(i,j)$ denote the polygon with vertices~$i$, $i+1$,
$i+2$, $\ldots$, $j-2$, $j-1$, $j$, $j-1$, $j-2$, $\ldots$, $i+2$,
$i+1$, $i$ in order, where the indices are taken modulo~$n$. Imagine
cycles of zero length around each point. Repeatedly introduce a new
cycle into~$\Pants'$ that is obtained by merging the two cycles adjacent along the
tour enclosing the fewest number of points. Each cycle $C(i,j)$ is obtained by merging two cycles $C(i,k)$ and $C(k+1,j)$ by doubling the edge between vertices~$k$ and~$k+1$. We ensure that each edge in the tour~$T$ appears exactly twice in at
most $\ceil{\log n}$ cycles in the pants decomposition
$\Pants$. Hence, $\abs{\Pants} \le 2 \ceil{\log n} \abs{T}$.

It is well-known~\cite{papadimitriou82combinatorial} how to obtain a
$\nicefrac{3}{2}$-approximate shortest Euclidean TSP of the point set
using Christofides' algorithm in $O(n^3)$ time; the minimum spanning
tree of the~$n$ points can be used to obtain a $2$-approximation in
$O(n \log n)$ time.  An approximate TSP tour obtained by either of
these algorithms gives us a non-crossing pants decomposition of length
$O(\log n)$ times the optimum.

\section{PTAS}
\label{sec:ptas}

Let $\e > 0$ be an arbitrary constant.  To construct in polynomial
time a $(1+\e)$-approximation to the shortest non-crossing pants
decomposition, we modify the PTAS for Euclidean TSP tour due to
Mitchell~\cite{mitchell99guillotine,mitchell97gridrounded}.  Our
algorithm is more complicated because a pants decomposition consists
of $\Theta(n)$ cycles instead of just one cycle as in a TSP tour.  The
PTAS is a dynamic programming algorithm where each subproblem is a
rectangular region of the plane and two adjacent subproblems interact
only through $O(1)$ grid points or \emph{portals}.

Let $m \ge 2$ be an integer and let $M=m(m-1)$.  Let~$B$ denote the
axis-aligned bounding box of the point set~$P$. Imagine a shortest
pants decomposition~$\Pants^*$.  Mitchell~\cite{mitchell97gridrounded}
has shown that there exists a \emph{favorable cut}, i.e., a horizontal
or vertical line~$l$, which partitions~$B$ into two smaller boxes that
can be recursively subdivided using favorable cuts.  The recursion
stops when a box is empty of points of~$P$. Just like Mitchell, we
introduce a segment of~$l$, called a \emph{bridge}, and $O(M)$ grid points on
$l$.  Mitchell has shown that the total length of the additional
subsegments is at most $\frac{\root{2}}{m}$ times the length of
$\Pants^*$.

Let~$R$, a rectangle, be the boundary of an arbitrary box~$Q$ during
the recursive subdivision.  Intuitively, we can ``bend'' the cycles of
$\Pants^*$ to make each cycle that crosses~$R$ transversely do so
only at one of the portals (grid points) and possibly use subsegments of
the bridges on the four sides of~$R$, without increasing the length of
the pants decomposition by too much.

To construct a short pants decomposition~$\Pants$, our PTAS solves
subproblems of the following form.  We are given a rectangle~$R$ whose
sides are defined by two horizontal and two vertical favorable
cuts. We are given two integers~$n_i$ and~$n_t$, both in the range
from~$0$ through $n-1$, of the number of cycles of~$\Pants$ that are
inside~$R$ and that intersect~$R$ transversely, respectively. Each of
the~$n_i$ cycles inside~$R$ intersects~$R$ tangentially and an even
number of times, and each of the~$n_t$ cycles intersects~$R$
transversely and an odd number of times. Let $n_R = n_i + n_t$.  We
are given the pattern in which the~$n_R$ cycles intersect~$R$ at the
$O(M)$ grid points on the sides of~$R$.  There are $O(n^{O(M)})$
possible ways for the~$n_R$ cycles to intersect the sides of~$R$.

It remains to account for the fact that cycles in~$\Pants$ that
intersect~$R$ tangentially can traverse subsegments of the cuts
bounding~$R$.  We observe that every point in the plane lies on at most two
independent cycles of~$\Pants$.  Let~$p$ be an arbitrary point in the
plane. Let~$C_p$ denote the subset of cycles in~$\Pants$ that pass
through~$p$. If $\abs{C_p} > 2$, then there exist three cycles $C_1$,
$C_2$, and~$C_3$ in~$C_p$ such that both~$C_1$ and~$C_2$ are
inside~$C_3$.  Let~$C$ be an outermost (minimum depth) cycle that
traverses a subsegment~$ab$ of some cut. Each subsegment is shared by
at most two independent cycles.  Therefore, we count the length of
$ab$ at most twice when counting the total length of all subsegments
of cuts traversed.

The dynamic programming algorithm proceeds as follows. For a rectangle
$R$ intersected by $n_R=n_i+n_t$ cycles, we try each of the $O(n)$
favorable cuts that partition~$R$ into two smaller rectangles, $A$ and~$B$.
We try each of the $O(n_R^{O(M)})$ possible ways that the~$n_R$ cycles
can intersect the cut transversely, making sure that the pattern in
which the cycles intersects the cut is consistent with the pattern in
which they intersect~$R$.  Some of the~$n_i$ cycles that belong inside
$R$ may belong inside~$A$ and some others inside~$B$; we try the
$O(n_i)$ possible ways to allocate a subset of the~$n_i$ cycles to~$A$
and the remaining to~$B$. We optimize over the $O(n)$ cuts and
$O(n^{O(M)})$ intersection patterns to solve the subproblem~$R$
optimally.  In the base case, if~$R$ has no points of~$P$ in its
interior, then the subproblem has only $O(M)$ size, which is a
constant, and is solved by brute force.  Since there are $O(n^{O(M)})$
different subproblems and each subproblem takes $O(n^{O(M)})$ time,
the total running time is $O(n^{O(M)})$.

The length of the pants decomposition~$\Pants$ obtained by the dynamic
programming algorithm is $O\paren{1+\frac{2\root{2}}{m}}$ times that
of a shortest pants decomposition. To obtain the desired approximation
factor, we choose $m \ge 2\root{2}/\e$.

To reiterate, the major differences between our PTAS and that of
Mitchell~\cite{mitchell99guillotine} are the following. (i)~Each of
our $n-1$ cycles crosses transversely the boundary of a rectangular
subproblem in one of $O(M)$ different ways.  Therefore, we have
$O(n^{O(M)})$ times as many subproblems to solve as in the
TSP. (ii)~Each of the $O(M)$ grid points on the boundary of a
rectangular subproblem may lie on any of the $n-1$ cycles in a pants
decomposition. Therefore, the additional information associated with
each subproblem is more than of constant size; the amount of
information associated with each subproblem is $O(n^{O(M)})$.
However, the total running time is still polynomial in~$n$.

\section{Points on a line}
\label{sec:collinear}

Let~$P$ be a set of points on a line, without loss of
generality on the $x$-axis.  Order the~$n$
points from left to right.  Let~$x_i$ denote the $x$-coordinate of the
$i$th point. For every $1 \le i \le j \le n$, let $(i,j)$ denote the
$j-i+1$ consecutive points numbered from~$i$ through~$j$.

We prove next that a shortest non-crossing pants decomposition of
$\Surf=\thePlane \setminus P$ must consist of convex cycles only.
Hence, if~$\Pants^*$ is a shortest non-crossing pants decomposition of
$\Surf$, then there exists a~$k$ in the range $1 \le k < n$ such that
$\Pants^*$ consists of a shortest pants decomposition of $(i,k)$ and a
shortest pants decomposition of $(k+1,n)$ together with an outermost
cycle of length $2(x_n-x_1)$ enclosing all~$n$ points.

Suppose to the contrary that there is a cycle~$C$ in~$\Pants^*$ that
contains a non-contiguous subset of points in~$P$.  Without loss of
generality, we assume that~$C$ is a minimal one in the sense that both
its legs, $C_1$ and~$C_2$, contains contiguous subsets of
points. Without loss of generality, assume~$C_1$ is to the left of
$C_2$; thus, $C_1$ is the \emph{left leg} and~$C_2$ is the \emph{right
leg} of~$C$.

Let $X \subseteq P$ be the set of all points between~$C_1$ and
$C_2$. Let $x \in X$ be arbitrary. Let ${\cal D}_x = \{D_1, D_2, D_3,
\ldots, D_i, \ldots \}$ be the set of cycles in~$\Pants^*$ containing
$x$ such that~$D_{i+1}$ contains~$D_i$.  Let~$i$ be the smallest index
such that~$D_i$ contains some point of $P \setminus X$. There are two
cases to consider: (i)~$D_i$ contains~$C$, (ii)~$D_i$ does not contain
$C$.

If~$D_i$ contains~$C$, then we construct another cycle~$E$ enclosing
$C_1$ and~$D_{i-1}$ making~$E$ the left leg of~$C$ (instead of
$C_1$). Delete~$D_i$.

Otherwise, if~$D_i$ does not contain~$C$, then either (a)~$D_i$
contains points only to the left of~$C_2$ or (b)~$D_i$ contains points
only to the right of~$C_1$. In the former case, $D_{i-1}$ is the
\emph{right leg} of~$D_i$. We swap~$C_1$ and~$D_{i-1}$ so that~$C_1$
is the new right leg of~$D_i$ and~$D_{i-1}$ is the new left leg of
$C$.  In the latter case, $D_{i-1}$ is the \emph{left leg} of
$D_i$. We swap~$C_2$ and~$D_{i-1}$ so that~$C_2$ is the new left leg
of~$D_i$ and~$D_{i-1}$ is the new right leg of~$C$.

In either case, we obtain a pants decomposition with total length
smaller than~$\Pants^*$, which is a contradiction.

Let $c(i,j)$ denote the cost of a shortest pants decomposition of
$(i,j)$.  We have just proved that $c(i,j)$ satisfies the following
recurrence for every $1 \le i \le j \le n$:
\begin{equation}
  c(i,j) = 2 (x_j-x_i) + \min_{i \le k < j} \paren{c(i,k) + c(k+1,j)}
\label{eqn:collinear:recurrence}
\end{equation}
where $c(i,i) = 0$.  A shortest pants decomposition of $(i,j)$ can be
computed by choosing the appropriate value of~$k$ in the range $i \le
k < j$, computing the optimum pants decompositions of $(i,k)$ and
$(k+1,j)$, and introducing a non-crossing cycle of length $2 (x_j-x_i)$
enclosing the points $(i,j)$. The straightforward dynamic programming
algorithm computes $c(1,n)$ in $O(n^3)$ time.

We show how the running time of the dynamic programming algorithm can
be improved by a linear factor using Yao's
speedup~\cite{yao82speedup}. Let $w(i,j) = x_j-x_i$. The function
$w()$ is \emph{monotone}, i.e., $w(i,j) \le w(k,l)$ whenever $(i,j)
\subseteq (k,l)$, and satisfies the following \emph{concave quadrangle
inequality}~\cite{yao82speedup}:
\[
  \forall \, i \le i' \le j \le j':
  w(i,j) + w(i',j') \le w(i',j) + w(i,j')
\]
In fact, the above equation is satisfied with equality because
$w(i,j) + w(i',j')
= (x_j-x_i) + (x_j'-x_i')
= w(i',j) + w(i,j')$.
Let $c_k(i,j)$ denote $w(i,j) + c(i,k) + c(k+1,j)$. Let $K(i,j)$
denote the maximum~$k$ for which $c(i,j) = c_k(i,j)$.  The following
claims are almost identical to those in the context of optimum binary
search trees proved by
Mehlhorn~\cite{mehlhorn84datastructures,thite01msthesis}:
\begin{enumerate}
\item
  The function $c(i,j)$ also satisfies the concave quadrangle inequality, i.e.,
\[
  \forall \, i \le i' \le j \le j':\,
  c(i,j) + c(i',j') \le c(i',j) + c(i,j')
\]
\item
  $K(i,j-1) \le K(i,j) \le K(i+1,j)$
\end{enumerate}
We compute $c(i,j)$ by diagonals, in order of increasing value of
$j-i$.  For each fixed difference~$d$, we compute $c(i,j)$ where $j =
i+d$; we compute $c_k(i,j)$ for
$k$ in the range $K(i,j-1) \le k \le K(i+1,j)$. The cost of computing
all entries on the $d$th diagonal is
\begin{align*}
&\sum_{i=1}^{n-d} K(i+1,j) - K(i,j-1) + 1\\
&= K(n-d+1,n+1) - K(1,d) + n - d\\
&\le (n+1) - 1 + n - d\\
&< 2n
\end{align*}
Since~$d$ ranges from~$0$ through~$n-1$, the total running time is
$O(n^2)$.

\section{Box decomposition}
\label{sec:box}

A \emph{box decomposition} of~$\Surf$ is a pants decomposition
$\Pants$ in which each cycle in~$\Pants$ is an axis-aligned rectangle.
Observe that any two axis-aligned rectangles can be separated from
each other by either a horizontal or a vertical line.

Let~$x_1$ through~$x_n$ denote the $x$-coordinates of the~$n$ points
in increasing order, and let~$y_1$ through~$y_n$ denote the
$y$-coordinates of the~$n$ points in increasing order.  Let $h(i,j) =
2(x_j-x_i)$ and let $v(i,j) = 2(y_j-y_i)$. Let $w(i_1,i_2,j_1,j_2) =
h(i_1,i_2) + v(j_1,j_2)$; then, $w(i_1,i_2,j_1,j_2)$ is the perimeter
of the axis-aligned box whose sides have $x$-coordinates~$x_{i_1}$ and
$x_{i_2}$ and $y$-coordinates~$y_{j_1}$ and~$y_{j_2}$. The function
$w()$ is \emph{monotone} and satisfies the \emph{concave quadrangle
inequality}.

The rest of the proof is similar to the case for points on a line.  Let
$c(i_1,i_2,j_1,j_2)$ denote the cost of a shortest pants decomposition
of the points in the box $[x_{i_1},x_{i_2}] \times
[y_{j_1},y_{j_2}]$. The cost $c(i_1,i_2,j_1,j_2)$ obeys a recurrence
that is a two-dimensional generalization of
Equation~\ref{eqn:collinear:recurrence}. Similar to the dynamic
programming algorithm for points on a line, we compute
$c(i_1,i_2,j_1,j_2)$ by diagonals, in order of increasing value of
$\max\{i_2-i_1,j_2-j-1\}$.  For each pair of differences~$d_1$ and
$d_2$, we compute $c(i_1,i_2,j_1,j_2)$ where $i_2=i_1+d_1$ and $j_2 =
j_1+d_2$.  The cost of computing all entries on the diagonals defined
by $(d_1,d_2)$ is $O(n^2)$; since there are $O(n^2)$ such pairs
$(d_1,d_2)$, the total running time is $O(n^4)$.

\section{Work in progress}
\label{sec:open}

We mention some very interesting open questions that we are currently
investigating.

Is it NP-hard to determine, for an arbitrary~$L$, whether there exists
a non-crossing pants decomposition of the punctured plane of length at most
$L$? Is there a simple algorithm to compute an $O(1)$-approximate
shortest pants decomposition?

Are the cycles in a shortest (non-crossing) pants decomposition always
convex?  If not, how much longer than optimum is a convex pants
decomposition?

How efficiently can we compute a shortest pants decomposition of the
plane with different types of punctures, e.g., rectangular holes
instead of points?

How efficiently can we compute a shortest pants decomposition of other
2-manifolds, such as the torus minus a set of points?

\subsection*{Acknowledgments}

We thank members of the Algoritmiek group at TU/e, especially Mark de Berg, for
fruitful discussions.  Shripad Thite thanks the organizers
and participants of the Meeting on Optimization Problems in
Computational Topology at ENS, Paris, in November 2005, especially
\'Eric Colin de Verdi\`ere and Jeff Erickson.


\small
\bibliographystyle{abbrv}

\end{document}